\documentclass{article}

\RequirePackage[english]{babel}
\usepackage{graphicx}

\begin{document}

\begin{center}
{\Large \bf PROBLEMS OF STAR FORMATION THEORY AND PROSPECTS OF SUBMILLIMETER OBSERVATIONS}

{\large  D.~Z. Wiebe, M.~S. Kirsanova, B.~M. Shustov, \\ 
Ya.~N. Pavlyuchenkov}

{Institute of Astronomy, Russian Academy of Sciences, Moscow, Russia}
\end{center}

Astronomy Reports, in press

\section*{Abstract}
We consider current state of star formation theory and
requirements to observations in millimeter and submillimeter
ranges which are necessary for resolution 
of the most actual problems of the
physics of star formation. Two key features of
star-forming regions which define observational requirements to
their studies, are relatively low energy of processes that take
place there and smallness of corresponding spatial scales. This is
especially true for the objects in the latest stages of ``pre-stellar''
evolution, that is, hot cores, hyper- and ultracompact HII
regions, and protoplanetary disks. Angular
resolution, sensitivity, and spectral coverage 
in existing projects of ground-based and space
telescopes of submillimeter and millimeter range are not completely adequate
to necessary requirements. To obtain
detailed information on star-forming regions as well as on
individual protostars it is necessary to employ a space-based
interferometer.

\section{INTRODUCTION}

By  the end of twentieth century the overall picture
of star formation process in dense interstellar clouds
became fully formed. Contraction (collapse) of protostellar clumps
is initiated by gravitation and external pressure; thermal pressure, 
rotation of the clumps and magnetic field counteract contraction. Thus,
general scenario of star formation is defined by a complex interaction
of these factors. However, many fundamental problems of star formation 
remain unsolved.    For instance, the mechanisms that stimulate the collapse of 
low-mass prestellar cores remain unclear. Even more enigmatic is formation of
massive stars. Apparently, more complex processes, like competitive accretion and merger of protostellar 
fragments are involved in the latter process in addition to spherical and disc accretion.

\begin{table*}[p!]
\caption{Atomic and molecular lines in the 10~mkm to 2~cm range which are used for the study of star formation regions.} 
\begin{tabular}{|l|c|l|}
\hline 
Molecule & Lines & Radiation mechanism \& Specific problem \\  
\hline
 H$_{2}$O & 22~GHz  & Maser emission, Studies of protostellar discs, astrometry \\
CH$_{3}$OH & 25~GHz & Maser and thermal emission, Astrometry, determination of \\
           &        & physical conditions in protoplanetary \\
           &        &discs and regions of massive star formation \\
NH$_{3}$ & 24~GHz & Thermal emission, Tracer of physical conditions \\
           &        &in the dense gas, especially in the late stages \\
           &        &of the evolution of prestellar cores \\
CS & 49~GHz  &  A tracer of physical conditions in the dense gas\\
   & 98~GHz  & \\
   & 147~GHz & \\
   & 244~GHz & \\
HCN & 89~GHz & A tracer of physical conditions in the dense gas \\
    & 266~GHz & \\
HCO$^{+}$ & 89~GHz & Tracer of physical conditions \\ 
  & & in the dense gas, including ionization degree \\
    & 268~GHz & \\
HNC & 91~GHz  &  A tracer of physical conditions in the dense gas \\
    & 272~GHz & \\
N$_{2}$H$^{+}$ & 93~GHz & Tracer of physical conditions \\ 
               &        & in the dense gas, especially in the late stages \\
      &  & of the evolution of prestellar cores and \\
      &  & in the regions of massive star formation \\
CO & 115~GHz & The main tracer of the presence of diffuse molecular gas\\
   & 230~GHz &\\
H$_{2}$CO & 140~GHz &  Tracer of physical conditions in the dense gas\\
NO        & 150~GHz &  Tracer of physical conditions in the dense gas \\
          & 250~GHz &\\
H$_{2}$D$^{+}$ & 372~GHz & Tracer of physical conditions \\ 
               &         & in the dense gas, especially, of the \\
               &         & kinematics of the central regions of prestellar cores \\
C & 492~GHz & Tracer of physical conditions in diffuse gas,  \\
  &         & PDR-regions, ultracompact HII regions \\
  & 809~GHz & \\
C$^{+}$ & 1.9 THz & A tracer of physical conditions in diffuse gas, \\
        & &PDR-regions, ultracompact HII regions\\
Si$^{+}$ &  8.6 THz & A tracer of physical conditions in the protoplanetary discs\\
H$_{2}$  & 10.7 THz & A tracer of physical conditions in the protoplanetary discs\\
Fe$^{+}$ & 11.5 THz & A tracer of physical conditions in the protoplanetary discs\\
S & 12.0 THz & A tracer of physical conditions in the protoplanetary discs\\
Fe & 12.5 THz & A tracer of physical conditions in the protoplanetary discs\\[1mm]
\hline
\end{tabular}
\end{table*}

\begin{table*}[t!]
\caption{Requirements to the angular resolution in observations of different stages of star formation}

\begin{tabular}{l|c|c|c}
\hline
 Stage & Typical  & Typical  & Angular resolution\\
 & distance & scale & (10 diagrams per object) \\
\hline
Prestellar cores  & 140 pc & 0.1 pc\phantom{0} & $15^{\prime\prime}$\phantom{0,} \\
Hot cores, UCHII$^*$ & 2--4 kpc and more & 0.1 pc\phantom{0} & ${<}1^{\prime\prime}$\phantom{00} \\
protostellar objects & 400 pc & 0.01 pc & \phantom{0}$0.5^{\prime\prime}$ \\
Outer regions of the discs & 100 pc & 1000 a.u. & $1^{\prime\prime}$\phantom{,} \\
Inner regions of the discs, & 100 pc & \phantom{0}100 a.u. & \phantom{0}$0.1^{\prime\prime}$ \\
brown dwarfs discs &&&\\
\hline
\end{tabular}

\end{table*}

\vspace*{5pt}
For the later evolutionary stages, when the planetary systems form, the
number of unsolved problems is none the less.
There are still no  unique solutions for the problems of the nature 
of angular momentum transfer in the protoplanetary discs,
their physical and chemical structure, the 
role of mixing in the  formation of chemical and mineralogical 
composition of protoplanetary matter (including  protosolar nebula).

These uncertainties are related, first, to the low energetics of 
the transformation of gas into stars, especially in its initial stages, 
which makes impossible studies of this process in the visual waverange. 
For instance, in the prestellar cores the temperature does not exceed 
10~K and it is only slightly higher at the periphery of protostellar 
objects, protoplanetary  and more evolved debris discs.
Therefore, a significant fraction of their radiation is emitted in submillimeter and millimeter waveranges.
However, these waveranges are very informative "--- the spectral range from 100~mkm 
to 20~mm contains thousands of lines of many dozens of interstellar molecules
(table~1) which, in the absence of observed emission of
molecular hydrogen, are the only source of information on composition,
temperature, and kinematics of molecular clouds and star formation regions.
Second, prestellar and, especially, protostellar objects are rather compact, but
at the same time their structure is complex and its study demands high angular resolution (table~2).
Just for this reason there is world-wide growing  interest to construction of
sensitive detectors for millimeter and submillimeter wavebands, including
interferometers.

Of course, the main expectations are related to the submillimeter 
interferometric system ALMA
(located in the Atacama desert in Chile) which will consist of
several tens of 12-m antennas with the maximum distance between them of 12~km~[1].
A somewhat lower scale project works successfully already. It is submillimeter
interferometer SMA~[2] in Hawaii (USA) which consists of eight 
6-m antennas. However, capabilities of the best of existing and planned ground-based 
instruments for the studies in millimeter and submillimeter ranges are limited by
disturbances created by the Earth atmosphere.
In the submillimeter range there are only several transparent windows 
in the Earth atmosphere, with transmittance factor that does not exceed  
\vspace*{5pt}
60$\%$ even in the Earth regions with the best astroclimate \footnote{http://www.eso.org/projects/alma/specifications/
FreqBands.html.}.
To some extent this problem will be solved with the launch of the space-born
infrared telescope ``Herschel'', but its spatial resolution will be limited 
by its
relatively small 3.5-m dish.
 More, in the long wavelength band ``Herschel''
will be sensitive up to 670!mkm only (ALMA will be sensitive up
to 1~cm). Therefore, this telescope Will be unable to observe
both the coldest clouds and many astrophysically interesting molecular lines,
including, for instance, the lines associated with low-level 
rotational transition of the CO molecule.

 A prerequisite to substantial increase in our knowledge of star formation
would be construction of an extra-atmospheric submillimeter telescope
which would conjoin sensitivity and spectral resolution with high angular resolution
 (which would mean possibility of its usage in the regime of an interferometer). 
It is by no means unimportant that location of the telescope in the space
would allow homogeneous observations both in northern and
southern    celestial
hemispheres. Space telescope ``Millimetron'' which is a part of the 
Russian Federal Space Programme may become such an instrument. 
In Russia, problems of star formation are studied in most of 
astronomical research centers (see, for instance, the collection of works edited by
Wiebe and Kirsanova~[3]). For this reason, high demand for observations 
by submillimeter telescope is expected from Russian scientists working
in this field of astrophysics. In the present paper we describe some of the problems
for which space-born submillimeter observatory will give crucially important
results and justify some of requirements to its parameters.

\section{THE STRUCTURE OF MOLECULAR CLOUDS, INITIAL CONDITIONS FOR STAR FORMATION}   

 Molecular clouds have complex structure which is associated
with chaotic motions in them. Most of modern researchers think that the structure and evolution of
molecular clouds, in particular, parameters of star formation in them   
(efficiency, mass function, kinematics of their cores) are defined
by turbulence~[4]. Apart of structural features, important role
of turbulence is indicated by different scale relations in the observed parameters of molecular
clouds (``velocity dispersion--size'', ``density--size''). However, 
parameters of turbulence "--- its outer scale and dissipation scale and sources 
of excitation  "--- are still unclear. For the studies of molecular gas
in our and other galaxies  the CO(1--0) lines are used traditionally, applying
so called X-factor, which is the ratio of column density   H$_{2}$ and integral
intensity of the  CO(1--0) line. However, CO(1--0) line is not always suitable
for solution of this problem. In some regions its emission is optically thick 
and, with increase of number density of molecular hydrogen,  X-factor loses
its  informativity. In other cases, relative abundance of CO molecules along the line of sight
is varied. As a result,  an uncertainty in the determination of the abundance of H$_{2}$ -molecules 
that exceeds and order of magnitude appears~[5, 6].

\begin{figure*}[t!]
\includegraphics{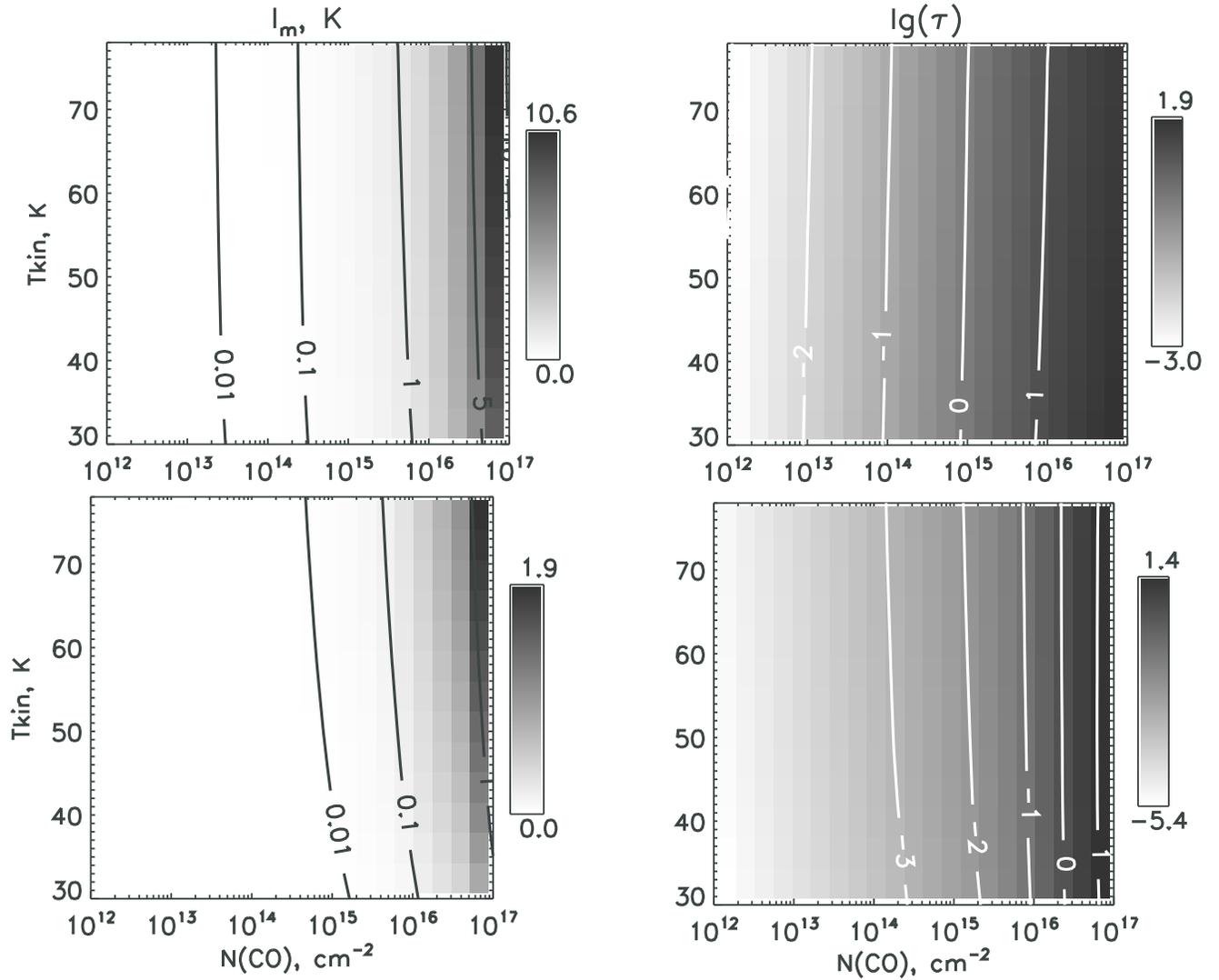}
\caption{Maximum brightness in units of $T_{\textrm{R}}$ (panels to the left)
and optical thickness (panels to the right) for CO(1--0)-lines (upper row) and
CO(4--3)-lines (lower row) as a function of kinematic temperature and column density 
of molecules in a homogeneous medium with density  
$n($H$_{2}) = 100$~cm$^{-3}$. 
Practically independent of temperature, optical thickness ~1 for CO(1--0) line is achieved at 
column density 
10$^{15}$~mol./cm$^{2}$, while for 
CO(4--3) line it is achieved at  
$3\times10^{16}$~mol/cm$^{2}$.
Thus,  CO(4--3) lines allow to penetrate ``deeper'' into dense regions of molecular clouds, but their brightness is by an 
order of magnitude lower than the brightness of   
CO(1--0) lines. \hfill}
\end{figure*}

Submillimeter range space interferometer with a wide bandwidth  will allow to solve this problem.
For the studies  of the structure and kinematics of the gas in diffuse clouds and less
dense 
regions of the nearest molecular clouds it would be useful to observe high-excitation lines
of the CO-molecule (Fig.~1) and its isotopologues, appended by observations of the lines
with visible hyper-fine splitting  C  and C$^{+}$. At present, such observations are successfully 
used for the studies of emitting regions in our and other galaxies.  However, to carry out
such observations for a more wide range of conditions, including ones in the cooler regions
of molecular clouds and at larger distances from the Sun, a larger sensitivity is necessary 
(up to several mJ for observations both in continuum and in the lines).
Simultaneous observations of CO, C, and C$^{+}$ in the clouds that do not 
show star formation activity
(for instance, in Maddalena cloud~[7]), as well as in the clouds with reduced abundance of molecular hydrogen,
allow tracing CO formation process and  measurement of  the ratios
 $^{12}$C/$^{13}$C and  $^{12}$CO/$^{13}$CO  for 
a large range of galactocentric distances, 
determination of the nature of the penetration of UV-radiation 
into molecular clouds,  derivation of the dependence of X-factor on external conditions.
The studies of CO in the submillimeter range may be appended by observations of
the CO transitions 
 in the ultraviolet range~[6]  by the space-born observatory 
``Spectrum -UV'' (also known as ``World Space Observatory "--- Ultraviolet").

High sensitivity will also allow  mapping   the thermal emission of the dust in these
regions and  studying the structure of 
magnetic field in them, by observations of dust emission by a wide band polarimeter. 
At present, ground-based instruments allow determination of 
 polarization of dust emission in 
dense clouds only. The structure of magnetic field recovered from these polarimetric
observations of the thermal dust emission does not correlate with the structure of
magnetic field in the surrounding less dense matter recovered from observations of the
thermal molecular lines and stellar radiation passing through the matter. The latter 
 is also polarized
due to interaction with the ensemble of dust particles aligned by magnetic field. 
Polarimetric observations of dust emission in the rarefied regions of molecular clouds 
will allow to resolve this controversy.

 There exists still unresolved problem of relation between mass function of prestellar clumps
and stellar initial mass function. Existing millimeter and submillimeter range instruments 
allow  already to map formation  regions of high- and low-mass stars~[8, 9].  For large enough viewing field of
the bolometers matrix (not less than 10$^\prime$ and, desirably, up to
30$^\prime$), a sensitive space-born submillimeter telescope would allow to carry out an in-depth
mapping of star formation regions and to study mass function of dense clumps, encompassing
not only low-mass stars but also sub-stellar objects (proto-brown-dwarfs, planetars, i. e. , isolated
objects of planetary mass). High sensitivity of the instrument will allow to study
star formation regions at large galactocentric distances and reveal possible 
connection of the parameters of the mass function of prestellar objects
(of so called initial mass function "--- IMF) with the gradient of 
chemical composition in the Galaxy and with its structure. For instance, observations of 
objects at different Galactic longitudes would allow to reveal possible variations of the parameters of 
IMF (in particular, relative distribution of formation regions of high- and
low-mass stars) with position of molecular clouds respective to the Galactic 
spiral wave.

\section{PRESTELLAR CORES}

One of the main sources of information on the physical conditions in the protostellar
objects are their spectra in a wide range of wavelengths (spectral energy distribution, SED). 
In essence, just SED became the basis for of the currently accepted system of 
classification of these objects.
Note, for class  ${-}1$ objects (starless cores) and class 0 ones (objects in
the very initial evolutionary stages in which central IR-source is present) 
the maximum of SED is located just in the submillimeter and millimeter ranges.
Location of the telescope in the cosmic space  will allow for the first time  
to construct SED of starless cores without discontinuities caused by existence
of atmospheric transparency windows. Detailed shape of the spectrum will
allow to answer at least two important questions. The first of them is related to
the evolutionary state of a specific core. Up to now, the cores were classified as
starless or protostellar ones on the base of presence or absence of
the central radiation source. Relative number of of starless
and prestellar serves as the basis for the estimate of the lifetimes in  respective
evolutionary stages. Though, for instance, observations of the 
starless core  L1014 by space-born infrared telescope ``Spitzer'' revealed 
that this core has a weak compact inner source, discovered due to excess of
radiation at wavelengths shorter than 70~mkm~[10]. Theoretical spectrum of this system
has a rather complex shape~[10, Fig.~3], but its reliability is based on several
observational points only which are obtained by different instruments. 
Space submillimeter telescope will allow to discover compact inner sources (future
stars or substellar objects) with even lower luminosity and in earlier
evolutionary stages, and will provide significantly more detailed
and homogeneous spectra extending from far IR to millimeter range for
verification of theoretical models.  

The properties of dust grains in starless and protostellar objects still remain
an important unsolved problem. Since the shape of the
dependence of  absorption coefficient on the frequency of radiation changes with 
with the size of grains, the growth of the latter is reflexed in the slope of the long-wavelenght spectrum.
 Construction of detailed spectra will, in particular, to find
at which stage of  the evolution of a protostellar object coagulation of the 
grains which finally leads to formation of dust clumps --- the embryo of
future planets --- begins.

Observations of high-excitation lines in prestellar cores will bring no less information 
then their observations in the diffuse medium. Strong molecular lines which are traditionally used 
for the studies of dense clumps in the molecular clouds
are a bad tracer of physical conditions in their most dense regions where where
formation of a protoplanet as such have already occurs or already occured.
Like in the case of the diffuse medium, this is related to the effects of radiation transfer and
chemical evolution. Due to high abundance of CO-molecules in the envelope of
a prestellar core, even lines of isotopologue  C$^{18}$O are optically
thick in transitions (1--0) and (2--1). Therefore, most convenient seem to be
high-excitation  lines of this isotopologue~[11], as well as the lines of isotopologue    C$^{17}$O. This is especially actual for the study of 
``chemically young'' cores,  which are currently actively searched for.
It is beyond doubt that in the next few years a large number of such cores
will be discovered, but without a high-sensitivity instrument their large-scale
investigation will be impossible.  
 
\begin{figure}[t!]
\includegraphics[height=10.cm]{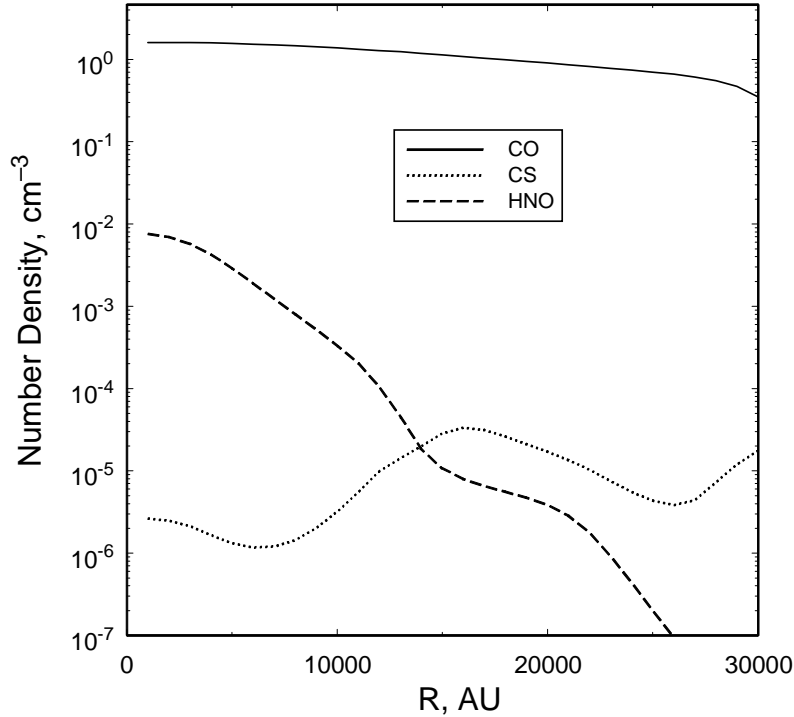}
\caption{Number density of some molecules in the model of starless core 
L1544 at the age of 2~Myr. The parameters of the model are as follows:
adhesion coefficient  "--- 0.3, intensity of UV-illumination in the units of interstellar
background "--- 0.2, 
the rate of ionization by cosmic rays "--- $\zeta = 10^{-17}$~s$^{-1}$. \hfill}
\end{figure}

In the more evolved cores, the effects of freeze-out are strong. Due to them the 
molecules that are usually observed in the central regions of prestellar cores
are bound in the ice mantles of dust grains. Since for the theory of star formation
kinematics of just these regions is  most interesting, one will be forced to use
for their study  low-abundance 
molecules (with relative abundance of molecular hydrogen $10^{-9}$ and less) and/or
new molecules (or transitions). In this respect, molecules containing nitrogen are
most promising, including not only traditional NH$_{3}$,
N$_{2}$H$^{+}$, HCN~[12], but also less known ones, like NO~[13]
and HNO.
Figure~2 shows theoretical radial
distributions of the absolute number densities of molecules CO, CS,
 and HNO in the starless  core L1544 computed by the model described in~[11]. 
It is seen that the number density of CO almost does not change 
along the profile of the core, while the number density of CS on average even 
slightly decreases to the center. Emission of these molecules is generated in the 
envelope of the core mainly and, therefore, it is a bad tracer of conditions in the central region of the core.
On the other hand, number density of HNO molecules in the envelope is minor, but
it substantially increases to the center. Also, as it was discovered recently,
it is convenient to use the lines of   H$_{2}$D$^{+}$ and other deuterized 
molecules for the studies of the kinematics of central regions of starless 
cores~[14]. These molecules also emit in the submillimeter range.

It is important to note that the width of spectral lines of starless cores usually
does not exceed several 100~m/s. Therefore,
for their observations one needs a high spectral resolution,
which enables the accuracy of determination of radial velocities of the order
of 10~m/s (not less than 50~m/s). For observations of nitrogen-rich molecules, 
the long-wave limit of high-resolution spectrometer must be not less than 15~mm.

\section{PROTOSTELLAR OBJECTS}

With a well-thought choice of molecules and transitions for the studies
of starless cores, it will be possible to obtain unique results even if space
telescope will be used in one-mirror regime.
However, investigations of more advanced stages of star formation
will require a significantly higher angular resolution which may be
achieved in the interferometer regime. Resolution of ``space antenna -- ground
antenna'' interferometer will allow to study the inner structure of protostellar 
objects at distances up to several kpc from the Sun. It will become possible,
for instance, to estimate directly from observations the scale of the region
in which the bipolar outflow is generated and to find whether it is due to the
wind of a young star in its immediate vicinity
or to the more powerful wind that encompasses protostellar   
disc  and contributes to the angular momentum loss. There will appear also a
possibility to study connection between outflows and continuing accretion 
onto a protostar at scales less than 1~a.u.

The detailed investigation of the kinematics and molecular composition
of the vicinities of young massive stars is necessary for solution of the problem
of their formation. Observations by space telescope in the
interferometric regime with high resolution will allow to study the structure of
distant regions of high-mass star formation,  to perform a comparative analysis 
of high- and low-mass protostellar objects.  Even now, observations of molecular lines from high-mass formation regions made by ground-based 
telescopes in the millimeter and submillimeter ranges revealed their
 complex structure~[15], but for quantitative interpretation of these
observations by chemo-dynamical models a significantly higher angular resolution is necessary.

Currently, quite actual are observations of immediate vicinities of protostars of
different masses. In particular,  the question about evolutionary connections between
hot cores, hypercompact regions of ionized hydrogen, and ultracompact regions
of ionized hydrogen remains unanswered, it is unknown whether they represent
subsequent stages of the region that 
surrounds a young massive star.   
Evolution of different fronts in the matter surrounding massive stellar
objects "--- shocks, ionization and dissociation (of H$_{2}$-molecule) fronts, evaporation
one "--- are of large interest, but due to 
insufficient sensitivity and angular resolution 
of modern instruments even the question about presence of dust in the vicinity of 
young stars (in the compact regions of ionized hydrogen)
remains unanswered.

Detailed observations of molecular composition of protostellar objects
with high angular resolution are necessary for the studies
of the details of evaporation of ice mantles of dust grains close to protostellar
objects. According to current notions, chemical reactions in these mantles 
result in formation of very complex molecules, including the
simplest organic ones. Observations of molecular spectra of the above mentioned 
regions will allow to explore the problem of possibility of nascence 
of organic molecules and their further propagation  through the cosmic space,
which is of extreme importance for  problem of the origin of life. 
In particular, it is necessary to learn whether these complex molecules desorb (volatilize)
from ice mantles without simultaneous destruction and what is their lifetime  before
destruction by short-wave stellar radiation.

\section{MASER SOURCES IN THE STAR FORMATION REGIONS}

Molecular maser emission is an irreplaceable tracer of the physical conditions and
processes in the star formation regions. It has a large potential for separation 
of different stages of evolution of young stellar objects.
This is due to high sensitivity of the parameters of maser emission to the parameters
of the medium in which it is generated. The main maser transitions which are observed in star
formation regions belong to the molecules of water  (H$_{2}$O), hydroxide (OH), and
methanol ((CH$_{3}$OH). Already at present Russian researchers are actively studying 
physical conditions in the interstellar medium using maser emission, in particular, 
that of methanol (see the papers published by V.~I.~Slysh group from Astro-Space center
of Lebedev Physical Institute   and by A.~M.~Sobolev group from Ural State University,
e.g.,~[16, 17]).
 Maser sources (spots) have  angular dimensions of fractions of an arcsecond 
(down to
a thousandth and less). Therefore, for their studies one needs interferometers. High
angular resolution of the ``ground--space'' interferometer and possibility of observations
in several spectral ranges will allow to study the most important questions 
of the theory of formation of masers:
\begin{itemize}
\item Which stages of the evolution of young stellar objects
are accompanied by appearance of maser emission?

\item What is the morphology of the regions of maser emission?

\item Where in space maser emission is generated respective to a young stellar object?
\end{itemize}

\begin{figure*}[t!]
\includegraphics[height=10.cm,angle=270]{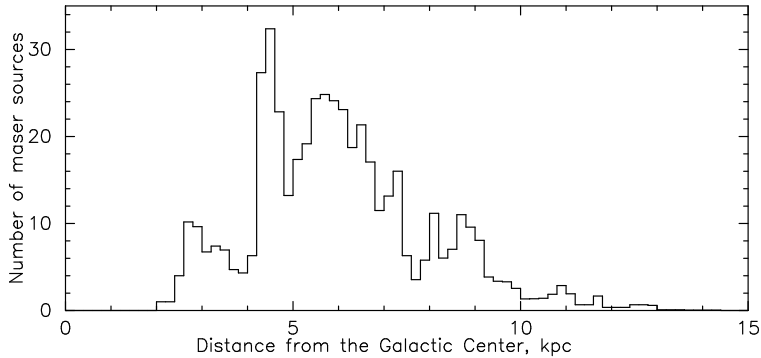}
\caption{Dependence of the class II methanol masers on the galactocentric 
distance. The diagram is provided by A.~M.~Sobolev and based on data 
from~[22]. \hfill} 
\end{figure*}

The most suitable for interferometric observations are class~I
water and methanol masers
(with frequencies  84, 95, and 104~GHz). In particular, they were discovered 
in a young stellar object IRAS 16547-4247~[18].

One more problem which may be solved by observations of masers is determination
of distances. The measurements of trigonometric parallaxes
of the sources of water and methanol maser emission allows to determine the distances
to the Galactic star formation regions avoiding all uncertainties related to
restricted capabilities of the kinematic method (uncertainty ``close-large distance''
for the sources from the inner galactic regions, necessity of usage of rotation curve).
The parallaxes of maser sources allow to determine distances to the sources up to
several kpc from the Sun, still unattainable for optical and infrared instruments. 
Measured distances will allow to determine positions of star formation regions in the Galaxy
and their concentration to the spiral arms and, hence, to find out the role of 
spiral structure in the star formation process. For instance,  the distance to 
the high-mass stars formation complex W3 in the Perseus spiral arm was measured
using parallaxes of the methanol maser sources. It is  $2\pm0.1$~kpc~[19, 20].
Determination of distances to the regions of star formation in distant outer Galactic
regions (galactocentric distances up to 15~kpc) will allow to construct a precise
rotation curve for the Galaxy.  The first steps in this study are made already for the star formation
region S269 located at 5~kpc from the Sun~[21]. A sensitive space submillimeter
interferometer
would significantly expand the possibilities for solution of this fundamental astronomical task.

Observations of masers also allow to learn how their number changes depending 
on galactocentric distance. In Fig.~3 we show distribution of class~II 
methanol masers in the galaxy according to Sobolev et al.~[22]. 
It is seen that the number of masers increases at galactocentric distances that
correspond to the positions of spiral arms. Observations by space instrument 
will allow to study Galactic distribution of the other type masers, in particular, 
to find out, whether their number depends not only on the presence of molecular gas, but also on its metallicity (in connection with existence of Galactic gradient
of chemical composition).

It is important to note that all above mentioned studies require a possibility of
interferometric spectral observations.

\section{CIRCUMSTELLAR DISKS}

The final stage of the evolution of a low-mass protostellar object is a T Tauri
star surrounded by gas-dust disk. Radiation of these disks is of large interest
for numerous reasons.

 \begin{itemize}
\item First, according to current notions, just in them planetary systems similar
to the solar one occurs.

\item Second, the studies of relatively close protoplanetary disks will allow 
to investigate a fundamental physical problem "--- the mechanisms of angular momentum transfer 
in the accretion discs.

\item Third, the processes of mixing associated with the transfer of angular momentum
in the protoplanetary discs may explain the peculiarities of chemical and isotopical
composition of the solar system.

\item Finally, in the inner regions of the discs dust grains retain ice mantles 
with complex molecular composition. The presence of the mantles influences both
the processes of grains coagulation (the growth of the planetesimals) and the
chemical composition of the embryos of planets which are born in this process.
\end{itemize}

Angular resolution of modern instruments is insufficient for full-scale
studies of protoplanetary discs.  Currently, interferometric studies do not allow to
obtain the images of objects of interest. When interferometric system
ALMA will be put into operation,  it will become possible to study the structure of
discs in general, but investigation of the details of the structure, in particular,
of the regions cleared of the matter by young planets (existence of which is a
signature of an important stage in the evolution of planetary systems) requires 
an instrument with a higher angular resolution.

The studies of SED in the inner regions of protoplanetary discs will be possible 
even in the one-mirror regime of the work of space telescope. Under condition
of high sensitivity it will allow to follow distribution of matter at large distances from the central star 
and to determine the parameters of the growth of grains to larger dimensions than 
it is possible by present-day instruments. 
Observations of different transitions and molecular isotopologues will allow to 
investigate the vertical structure of the discs, their chemical and thermal structure,
to study efficiency of mixing~[23], as well as possibility of layered accretion. 

A possibility for the similar studies of brown dwarfs discs will appear. At present,
the studies of brown dwarfs attract a large attention due to 
uncertainties  existing in the scenarios of their formation. An important stage in the 
finding out of their origin became investigations of gas-dust discs around them. 
As it is shown by observations, while in general discs of brown dwarfs and of 
low-mass stars have similar physical structure, the discs of the former are less dense. 
As a result, they are more transparent for ionizing radiation (for X-rays and for cosmic rays) 
and this determines another chemical structure. In particular, theoretical calculations~[24]
show that column densities for majority of observed molecules in the discs of brown dwarfs
are by 1 to 2 orders of magnitude lower than in more massive discs.
In addition, indirect data give an evidence that the discs of brown dwarfs are more compact~[25]. 
As a result, existing ground-based telescopes allowed  to obtain observational data not only
on molecular lines but also on thermal dust emission   
only for several such discs in the close vicinity of the Sun. It is quite possible
that the operation of high-sensitivity space telescope will not simply
expand our knowledge about brown dwarfs, but also will change our notions about
peculiarities of chemical reactions in protoplanetary discs in general.

\section{CONCLUSION}

Currently, several projects of millimeter and submillimeter range telescopes
are already implemented or developed.  The problems related to different
stages of star formation take an important place in their programmes. 
However, the sensitivity of ground-based instruments is  naturally limited
by atmospheric absorption. A submillimeter telescope in the space, even operating in the
one-mirror would allow to carry out extremely important investigations
both of close and distant  star formation regions (thanks to high sensitivity)
and will allow to solve a lot of enigmas of star formation not only in the
vicinity of the Sun but also in a more extended region of the Galaxy.
To some extent this goal will be achieved by ``Herschel'' project, but the angular
resolution of `Herschel'' will be not high enough to solve many
fundamental problems of star formation theory. In this respect, capabilities of
the ``Earth--Space'' interferometer will be unbeaten and will allow to obtain
highly valuable 
information 
on the structure of the star formation regions and protostellar (young stellar) objects
that is unavailable by other means.

\section{ACKNOWLEDGMENTS}

The authors acknowledge A.~M.~Sobolev and A.~B.~Ostrovskij for their comments on observations
of masers.   
This study was supported by  the Program of state support  of leading scientific schools (project code НШ-4820.2006.2).
DZW also acknowledges  President of the Russian Federation grant for 
support of young Doctors of Science (project code 
МД-4815.2006.2).

\end{document}